\documentclass[aps,prl,amsmath,amsfonts,amssymb,twocolumn]{revtex4}
\usepackage[latin1]{inputenc}
\usepackage{graphicx}
\usepackage[english]{babel}
\usepackage[usenames,dvipsnames]{color}

\usepackage{amsfonts}
\usepackage{amsmath}
\usepackage{amssymb}

\batchmode

\usepackage{color}

\usepackage{soul}

\newcommand{\beq}{\begin{equation}}
\newcommand{\eeq}{\end{equation}}

\newcommand{\ket}[1]{|#1\rangle}
\newcommand{\bra}[1]{\langle #1|}

\makeatletter
\@ifundefined{textcolor}{}
{%
 \definecolor{BLACK}{gray}{0}
 \definecolor{WHITE}{gray}{1}
 \definecolor{RED}{rgb}{1,0,0}
 \definecolor{GREEN}{rgb}{0,1,0}
 \definecolor{BLUE}{rgb}{0,0,1}
 \definecolor{CYAN}{cmyk}{1,0,0,0}
 \definecolor{MAGENTA}{cmyk}{0,1,0,0}
 \definecolor{YELLOW}{cmyk}{0,0,1,0}
 }

\begin{document}

\title{Asymmetric environment induces protein-like relaxation in spin chain}

\author{D. Valente
$^{1}$
}
\email{valente.daniel@gmail.com}

\author{T. Werlang
$^{1}$
}
\email{thiago.werlang80@gmail.com}

\affiliation{
$^{1}$ 
Instituto de F\'isica, Universidade Federal de Mato Grosso, CEP 78060-900, Cuiab\'a, MT, Brazil
}

\begin{abstract}
Proteins are aminoacid chains that diffusively fold or unfold depending on the thermal and chemical environmental conditions.
While sophisticated models account for detailed aspects of real proteins, finding traits that unify protein dynamics to general open chains relaxation is still challenging.
The principle of minimal frustration represents a key step towards this goal, revealing a fundamental link between proteins and spin glasses.
Here, we search for the emergence of protein-like relaxation in open spin chains by going beyond the validity domain of the minimal frustration principle as to focus on the role of system-environment interactions rather than on frustration in the system's Hamiltonian.
We find that strong asymmetries between the couplings of each spin to its immediate surroundings imply close similarities to protein folding and unfolding dynamics.
Namely, an asymmetric bath can (i) block the system from finding its minimum energy state, even in the complete absence of energetic frustration in the system's Hamiltonian, and (ii) excite the system resembling the well-known distinction between thermal and chemical denaturations.

\end{abstract}


\maketitle

Understanding relaxation mechanisms of diverse physical systems may allow us to unravel some of the physical principles underneath life-like behavior \cite{bialek,JE13,JE15}.
Proteins, for instance, are large sequences of aminoacids that relax from an unfolded to a final state.
In biologically functional proteins, final states are those folded in specific, compact geometries, called native states.
From the physics viewpoint, this protein relaxation dynamics is understood as diffusion in a rugged funneled energy landscape and the native states occupy the lowest energy subspace, i.e., the bottom of the funnel \cite{dill}.
Randomly chosen aminoacid sequences are unable to fold to its most compact, less energetic state.
Competing interactions between the aminoacids can impede, or frustrate, simultaneous minimization of all energetic contributions, creating many different states with nearly the same low energy, separated by large barriers.
Frustration can, thus, become responsible for ``blocking the system from finding a single well-isolated folded structure of minimum energy'' \cite{bialek}.
This raises the question of what physical principle may be behind those few aminoacid sequences that do fold properly and become biologically functional.
Theoretical and experimental evidence point towards the principle of minimal frustration \cite{wolynes, makha,yan}.
Frustration can also block magnetic systems from relaxing to their lowest energy levels, forming the so called spin glasses.
An emblematic model for spin glasses is the Sherrington-Kirkpatrick model \cite{SKM}, consisting of Ising spin chains with ferro- and antiferromagnetic couplings.
Interdisciplinary endeavors concerning relaxation and frustration date from the 80s, when the spin glass theory was first used to study protein folding and the principle of minimal frustration was coined \cite{wolynes}.

In the spirit of spin glasses, models from equilibrium statistical physics that successfully describe relaxation in protein folding, comprising G\=o and HP models \cite{bialek, bachmann}, are devised to solve for the kinetics and the thermodynamics of  a huge amount of degrees of freedom, where the computation of free energies is adequate.
Models that go a step further, as to address coarse-grained stochastic dynamics of proteins due to their environments, have benefited from the power of phenomenological approaches, such as Langevin dynamics and master equations with {\it ad hoc} friction and rate coefficients, to describe key aspects of protein folding and denaturation.
Remarkable examples are the description of protein denaturation dependence not only on temperature \cite{thirumalai1992, thirumalai1995, zhang2003, PREtorcini2008, EPLangelani2009, PRLkappler2019} but also on chemical concentration in the protein's environment, employing the molecular transfer model 
\cite{thirumalai2008, thirumalai2010, thirumalai2011,reviewJE, thirumalai2016}.

When we are interested in systems with a moderate number of degrees of freedom, especially when quantum effects play a role, explicitly accounting for microscopic system-reservoir interaction mechanisms turns out to be rewarding.
The so called system-plus-reservoir approach for open quantum systems \cite{caldeira,huelga, petruccione} sets a global Hamiltonian (allowing for quantization of both system and environment) that reproduces the stochastic dynamics of the quantum system of interest.
This approach provides a microscopic quantum theory for irreversible processes, such as the spontaneous emission of a photon by a single atom \cite{WW}, the decay of magnetic flux in a superconducting artificial atom \cite{CL} and quantum decoherence due to phonon baths in semiconducting artificial atoms \cite{besombes2001, dv14, mork2017}.
Under weak system-bath coupling regimes, the derived quantum master equations allow for the investigation not only of relaxation \cite{spohn,shishkov} but also of heat transport through quantum systems coupled to multiple reservoirs at different temperatures \cite{mahler}, including cases with intrachain ultrastrong couplings \cite{kosloff,werlang,werlang15}.

Here, we investigate how different distributions of system-reservoir couplings along a quantum spin chain alter its relaxation pathways, with the aim to disclose dynamical similarities between open quantum systems and protein folding and unfolding.
We derive a markovian quantum master equation valid for arbitrary spin-spin couplings that takes into account multiple independent reservoirs, all at the same temperature and weakly coupled to the spin chain.
We show general conditions for the bath to block a diffusive quantum dynamics from attaining its lowest energy levels, even in the absence of frustration in the system's Hamiltonian.
We also evidence how temperature and asymmetric couplings with independent reservoirs affect in distinct ways the stationary states of a spin chain. 
We illustrate this effect for an Ising chain of two spins.
Finally, we discuss how this difference entails striking resemblance with the well-known mismatch between thermal and chemical denaturation in proteins \cite{narayan}.

\begin{figure}[!htb]
\centering
\includegraphics[width=1.0\linewidth]{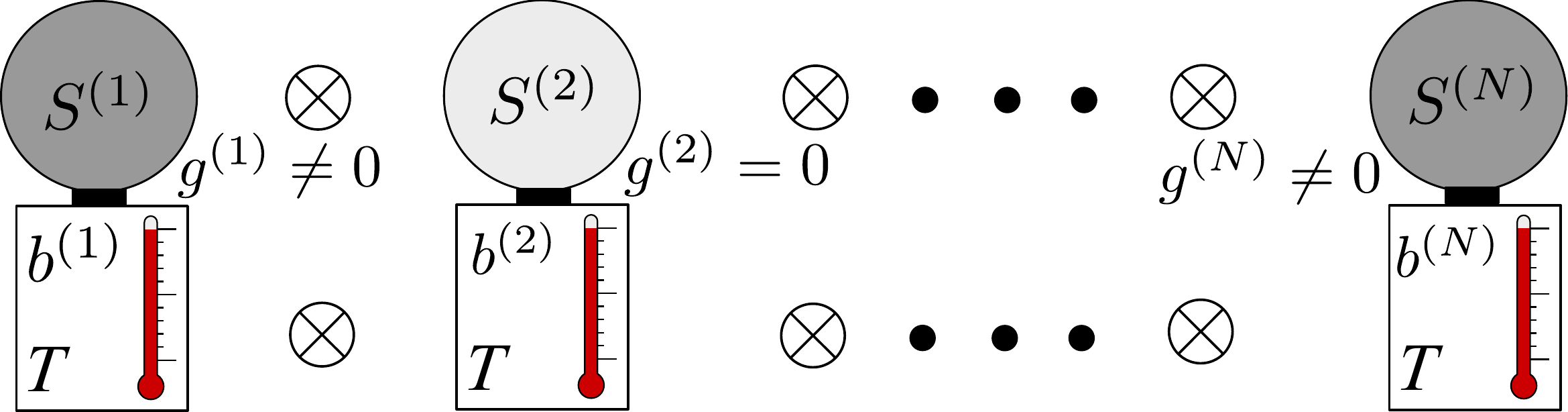}
\caption{(Color online)
System-reservoir couplings model.
The system is a chain of degrees of freedom $S^{(n)}$ locally coupled, with strengths $g^{(n)}$, to independent reservoir modes $b^{(n)}$, all at the same temperature $T$.
We are interested in the consequences of strongly asymmetric couplings $g^{(n)}$.
}
\label{fig1}
\end{figure}

We label as $H_S$ the Hamiltonian of a generic isolated quantum system.
The only assumption we need to make at this point is that we know its spectral decomposition, $H_S = \sum_{j=1}^d E_j \ket{j} \bra{j}$, where $d$ is the size of the Hilbert space.
Following the system-plus-reservoir approach, the Hamiltonian of the system coupled to its environment is set to
$H=H_S + H_{SR} + H_R$.
We model the reservoir Hamiltonian 
$H_R = \sum_{n=1}^{N} \sum_{k} \hbar\omega_{k}^{(n)} b_{k}^{(n) \dagger} b_{k}^{(n)}$ 
as a finite set of $N$ independent baths, each consisting of quantum harmonic modes $b_k$ of frequencies $\omega_k$ (that will be treated in the continuum limit, $\sum_k \rightarrow \int dk$).
Let us consider that each independent bath is locally coupled to a distinct degree of freedom $S^{(n)}$ of the system, as described by (see Fig.\ref{fig1})
\beq
H_{SR} 
= \sum_{n=1}^{N} S^{(n)} \otimes 
\sum_{k} \hbar g_k^{(n)} (b_{k}^{(n)\dagger} + b_{k}^{(n)}).
\label{HSR}
\eeq
In the case where the system is a spin-1/2 chain, for instance, $S^{(n)}$ may represent a Pauli operator.
For $N$ spins-1/2, we have that $d=2^N$.

We describe the state of our general quantum system by its density matrix, $\rho_S(t)$.
Our goal is to establish the quantum master equation governing the dynamics of $\rho_S(t)$.
To that end, we proceed by tracing out the environmental degrees of freedom from the complete quantum state evolved unitarily, $\rho_S(t) = \mbox{Tr}_R[U \rho(0) U^\dagger]$, where $U=\exp(-iHt/\hbar)$, from an initially uncorrelated global state $\rho(0)=\rho_S(0)\otimes \rho_R(0)$.
We choose a thermal equilibrium state for the reservoir at temperature $T$, so that $\rho_R(0)=\exp(-\beta H_R)/Z_R$, with $\beta=1/(k_BT)$, where $k_B$ is the Boltzmann's constant and $Z_R=\mbox{Tr}[\exp(-\beta H_R)]$ is the partition function.
We assume perturbative system-reservoir couplings up to second order. 
This allows us to characterize the couplings between each degree of freedom of the system to its local reservoir by the so called spectral function $J^{(n)}(\omega) = 2\pi \sum_k |g_k^{(n)}|^2 \delta(\omega-\omega_k^{(n)})$, which is well defined in the continuum limit, $\sum_k \rightarrow \int dk$.
These steps lead to the derivation of a markovian quantum master equation for the system density operator in the so called Lindblad form \cite{petruccione,huelga},
\beq
\partial_t \rho_S(t)=-(i/\hbar)[H_S,\rho_S(t)]+L[\rho_S(t)],
\label{ME}
\eeq
where $L[\rho_S(t)]$ supports the relaxation effects we wish to explore.
It reads
\begin{align}\label{L}
L[\rho_S] =& \sum_{n=1}^{N} \sum_{\omega>0}  
J^{(n)}(\omega) (1+\bar{n}_{\omega})
\Big[ 
A^{(n)}_\omega \rho_S  A^{(n)\dagger}_\omega
\nonumber\\
-&\frac{1}{2}\left\{ \rho_S, A^{(n)\dagger}_\omega A^{(n)}_\omega \right\}
\Big]
\\
+& J^{(n)}(\omega) \ \bar{n}_{\omega}
\Big[ 
A^{(n)\dagger}_\omega \rho_S  A^{(n)}_\omega 
-\frac{1}{2}\left\{ \rho_S, A^{(n)}_\omega A^{(n)\dagger}_\omega \right\}
\Big],\nonumber
\end{align}
where $\omega=\omega_{ij}=(E_j-E_i)/\hbar > 0$, the average number of excitations is $\bar{n}_{\omega}=[\exp(\beta\hbar\omega)-1]^{-1}$, as given by the Bose-Einstein distribution, and the jump operators are defined by
$
A^{(n)}_\omega = \sum_{i,j|\omega=\omega_{ij}} \ket{i}\bra{i}S^{(n)}\ket{j}\bra{j}.
$

The jump operators $A^{(n)}_\omega$ evidence the fundamental aspect retained by this microscopic approach, in that the bath acts locally, via  $S^{(n)}$, and affects globally, given that $\ket{j}$ is an eigenstate of the entire system.
This explains the potential of the model in Eq.(\ref{HSR}) to reveal rich relaxation phenomena.
The microscopic model, Eq.(\ref{L}), appropriately guarantees that the Gibbs state 
$\rho_S(\infty) = \exp(-\beta H_S)/Z_S$ 
is one (not necessarily unique \cite{spohn,shishkov}) steady-state solution for the open system dynamics.
This contrasts with phenomenological master equations, where the Lindbladian is derived under the assumption that each degree of freedom is decoupled from all the others.
In the cases we have checked, the phenomenological approach has led instead to a state  
$\propto \exp(-\beta \sum_\alpha H_\alpha)$, 
where $H_\alpha$ is the free Hamiltonian for the $\alpha$-th degree of freedom alone, i.e., $\sum_\alpha H_\alpha \neq H_S$.
As far as heat transport is concerned, the phenomenological approach may violate the second law of thermodynamics \cite{kosloff}. 
Besides being consistent with the second law, the microscopic model also reveals mechanisms not captured by the phenomenological approach, such as thermal rectification through Ising chains \cite{werlang} and heat transport induced by quantum pure-dephasing reservoirs \cite{werlang15}. 

We now establish our main results, namely, under what circumstances an asymmetric reservoir can 
(i) block the system from attaining its lowest energy levels and 
(ii) excite the system by distinct pathways as comparing the increase of temperature versus the coupling of a given site of the chain to its local reservoir.
As we try to make clear below, these general conditions have all the same origin, that is, the non-uniqueness of the steady-state $\rho_S(\infty)$ under asymmetric reservoir couplings along an open chain.
The relevant properties of $\rho_S(t)$ emerge when we rewrite it as a column vector $\vec{\rho}_S$.
We recast the master equation (\ref{ME}) in the form 
\beq
\partial_t{\vec{\rho}}_S = \Lambda \vec{\rho}_S,
\label{Lambda}
\eeq
where $\Lambda$ is a time-independent square matrix representing the transition rates between all the elements of $\rho_S(t)$.
We are interested in the typical case where the system's spectrum is nondegenerate.
A nondegenerate spectrum implies that the quantum coherences decouple from the populations \cite{petruccione}, allowing us to only focus on the latter, 
$(\vec{\rho}_S)_i= \bra{i}\rho_S\ket{i}$. 
Now $\Lambda_{ij}$ becomes the transition rate only between the energy eigenstates $\ket{j}\rightarrow\ket{i}$. 
Equation (\ref{Lambda}) then simply becomes the Pauli master equation.
We arrive here at the core of our results:
a block-diagonal $\Lambda$ implies a set of decoupled energy subspaces.
The stationary-state is not unique and it depends on the system's initial state, when $\Lambda$ is block-diagonal.
If the system has a finite probability of being initially excited with a certain energy outside the lowest-energy subspace, the relaxation pathway from the higher-energy to the lowest-energy subspace will be forbidden. 
The excited portion of the ensemble will be blocked from attaining the state of minimum energy, no matter how low the temperature is set.
This explains our result (i). 
It shows how an asymmetric bath can replace the role played by frustration in preventing open chains to achieve its minimum energy states.
The opposite pathway, that would lead to excitation, is also forbidden.
If the system starts trapped within a given low-energy subspace, there it will remain no matter how high the temperature is set.
The only mechanism that allows it to escape, as to achieve higher energy configurations, is by turning on the coupling between a degree of freedom and its local environment (breaking the block-diagonal structure of $\Lambda$, in our theory).
This distinction between exciting the chain by increasing the temperature $T$ in contrast to increasing a local coupling $J^{(n)}(\omega)$ justifies our result (ii), reminding us of the difference between thermal and chemical denaturations in proteins.

In order to establish how the distribution of couplings along the chain generates the desired decoupled subspaces (in other words, how $J^{(n)}(\omega)$ creates a block-diagonal $\Lambda$), we need an explicit form for $\Lambda$.
We add another simplifying condition, that the gaps $\omega_{ij}$ are also nondegenerate. 
We find that
$\Lambda_{ii} = \Gamma^{(0)}_i$, 
$\Lambda_{i<j} = \Gamma^{(D)}_{ij}$
and 
$\Lambda_{i>j} = \Gamma^{(G)}_{ij}$.
Following the Fermi's golden rule \cite{petruccione}, the off-diagonal elements here read
$\Gamma^{(D)}_{ij}=\sum_{n=1}^{N} J^{(n)}(\omega_{ij}) (1+\bar{n}_{\omega_{ij}}) |S_{ij}^{(n)}|^2$,
for the damping rates, and
$\Gamma^{(G)}_{ij}=\sum_{n=1}^{N} J^{(n)}(|\omega_{ij}|) \bar{n}_{|\omega_{ij}|}  |S_{ij}^{(n)}|^2$,
for the gain rates.
$S_{ij}^{(n)} = \bra{i}S^{(n)}\ket{j}$ 
are the matrix elements of the system's degrees of freedom in the energy basis.
Finally, the diagonal elements are given by
$
\Gamma^{(0)}_i = -\sum_{j=1}^{i-1} \Gamma^{(D)}_{ji} -\sum_{j=i+1}^{d} \Gamma^{(G)}_{ji}
$
for $1<i<d$,
$\Gamma^{(0)}_1 = -\sum_{j=2}^{d} \Gamma^{(G)}_{j1}$ and
$\Gamma^{(0)}_d = -\sum_{j=1}^{d-1} \Gamma^{(D)}_{jd}$.
Most importantly, rates $\Gamma^{(D)}_{ij}$ and $\Gamma^{(G)}_{ij}$ provide analytical expressions that show how the $N$ local and independent system-reservoir couplings, as quantified by $J^{(n)}(\omega)$, can cause a block-diagonal $\Lambda$, inducing the protein-like relaxation dynamics expressed in (i) and (ii) above.

\begin{figure}[!htb]
\centering
\includegraphics[width=1.0\linewidth]{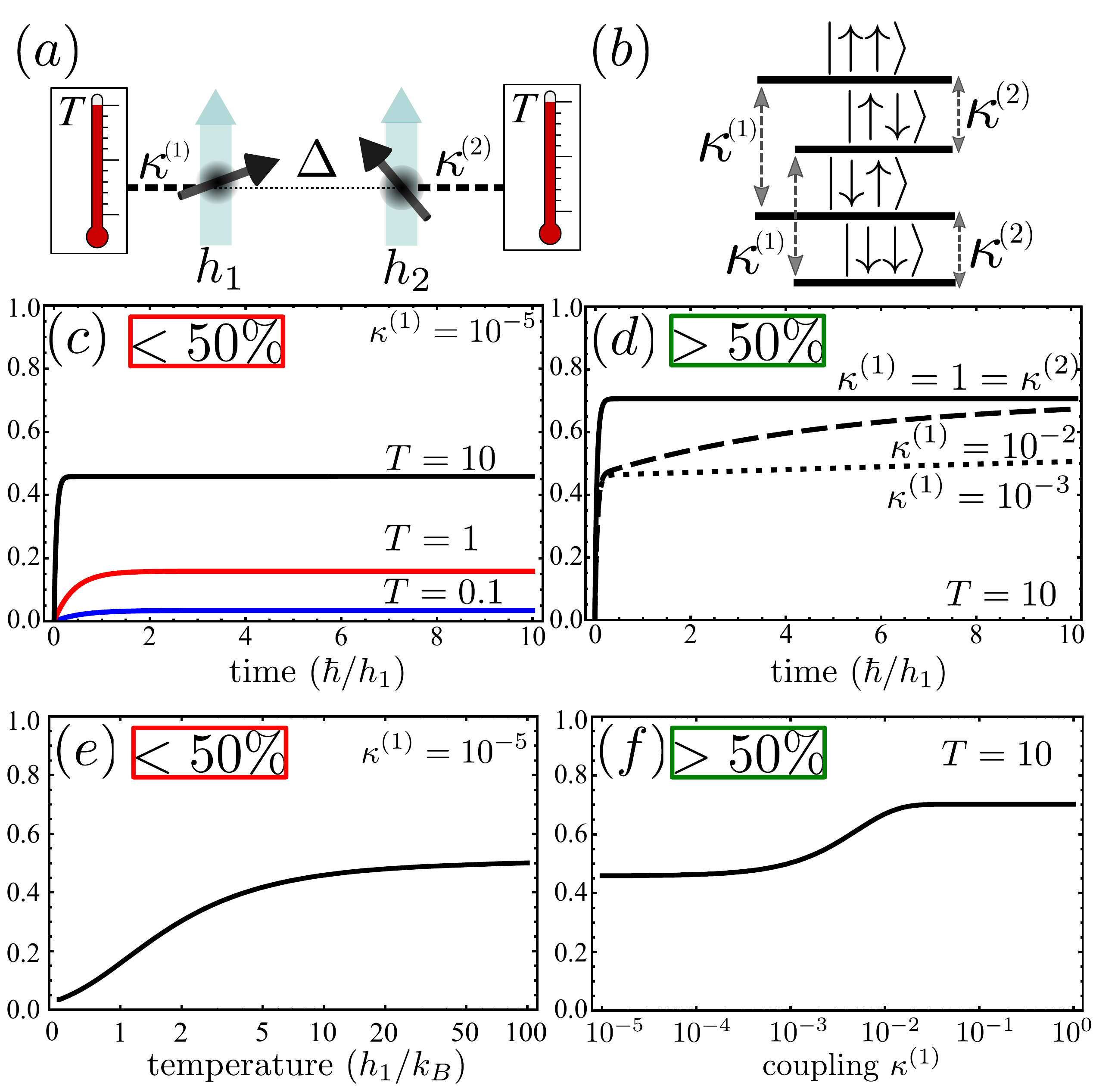}
\caption{(Color online)
Protein-like relaxation of an asymmetrically open Ising chain.
(a) Ising chain of two spins, black arrows represent $\sigma_z^{(n)}$.
(b) energy levels and relaxation pathways: vanishing $\kappa^{(1)}$ blocks low-energy subspace from high-energy subspace.
(c) and (d) Excitation probability $P_{\mathrm{exc}}(t) = 1-\bra{1}\rho_S(t)\ket{1}$ in time ($\hbar/h_1$ units).
(e) and (f) $P_{\mathrm{exc}}(t=10)$ with respect to $T$ ($h_1/k_B$ units) (e) and coupling $\kappa^{(1)}$ (f).
We set $h_2=h_1/2$, $\Delta=h_1/3$ and $\kappa^{(2)}=1$.
(c)-(f) protein-like denaturation: chemical-like excitations (varying $\kappa^{(1)}$) exceed the $50\%$ limit from thermal excitations at time $t=10$.
The higher chemical-like (f) as compared to thermal (e) excitation remarkably resemble the experimental results in \cite{narayan}.}
\label{fig2}
\end{figure}

Our next step is to illustrate our general statements (i) and (ii) with a well-known exactly solvable example.
We employ the Ising model, as inspired by \cite{SKM}, for a pair of spins, $N=2$.
See Fig.\ref{fig2}(a) below.
The system Hamiltonian is described by
$
H_S = h_1 \sigma_z^{(1)}+h_2 \sigma_z^{(2)}
- \Delta \sigma_z^{(1)} \sigma_z^{(2)}
$,
where $\sigma_z^{(n)}$ is the $z$-Pauli matrix of the $n$-th spin-1/2.
We choose $h_1>h_2>\Delta>0$.
The choice of parameter values shall guarantee the absence of energy frustration.
The chain is unfrustrated whenever the sum of the minimum energy of each term equals the minimum of the total energy.
We also make sure $H_S$ is nondegenerate and all transition frequencies $\omega_{ij}$ are unequal.
The energy eigenstates here are given by
$\ket{1}=\ket{\downarrow \downarrow}$,
$\ket{2}=\ket{\downarrow \uparrow}$,
$\ket{3}=\ket{\uparrow \downarrow}$ and
$\ket{4}=\ket{\uparrow \uparrow}$,
with eigenvalues
$E_1=-h_1-h_2-\Delta$,
$E_2=-h_1+h_2+\Delta$,
$E_3= h_1-h_2+\Delta$ and 
$E_4= h_1+h_2-\Delta$.
We are interested in energy-exchanging system-reservoir couplings, that satisfy $[H_S,S^{(n)}]\neq 0$.
We assume here that $S^{(n)} = \sigma_x^{(n)}$.
We finally choose an ohmic spectral function,
$
J^{(n)}(\omega) = \kappa^{(n)} \omega
$,
where $\kappa^{(n)}$ is a dimensionless parameter that we consider here as a free variable.

Figure \ref{fig2}(b) illustrates our result (i).
It shows the energy levels of our Ising chain.
Dashed arrows indicate the relaxation pathways induced by the two independent baths, characterized by 
$\kappa^{(1)}$ and $\kappa^{(2)}$.
Bath $(1)$ induces, via $\sigma_x^{(1)}$, transitions $\ket{1}\leftrightarrow \ket{3}$ and $\ket{2}\leftrightarrow \ket{4}$.
Bath $(2)$ induces, via $\sigma_x^{(2)}$, transitions $\ket{1}\leftrightarrow \ket{2}$ and $\ket{3}\leftrightarrow \ket{4}$.
If we make $\kappa^{(1)} = 0$, subspace $\left\{ \ket{1}, \ket{2} \right\}$ becomes decoupled from $\left\{\ket{3}, \ket{4} \right\}$.
The system starting its dynamics at the highest energy subspace, $\left\{\ket{3}, \ket{4} \right\}$, gets blocked from attaining the lowest energy subspace, $\left\{\ket{1}, \ket{2} \right\}$.

Figures \ref{fig2}(c)-(f) illustrate our result (ii).
A typical protein denaturation experiment follows the state of the system as a function of temperature at a given chemical concentration and compares it to the variation in the denaturant concentration at a constant temperature (see \cite{narayan}).
We follow a similar protocol.
In Figs.\ref{fig2}(c) and (d), we compute the excitation probability in time, defined here as $P_{\mathrm{exc}}(t) = 1-\bra{1}\rho_S(t)\ket{1}$.
To recall a protein-like denaturation dynamics, we start from our native-like state, $P_{\mathrm{exc}}(0)=0$.
Now we compare the two types of excitation processes in time ($\hbar/h_1$ units), i.e., the thermal versus the chemical-like.
In the thermal excitation process, we let $\kappa^{(1)}=10^{-5}$, $\kappa^{(2)}=1$ and obtain $P_{\mathrm{exc}}(t)$ at temperatures $T=0.1$ to $10$ ($h_1/k_B$ units).
We set $h_2=h_1/2$ and $\Delta=h_1/3$.
We see a saturation $P_{\mathrm{exc}}(t) \lesssim 50\%$ at high temperatures.
In the chemical-like excitation process, we keep the high temperature $T=10$ and vary the coupling $\kappa^{(1)}$ from $10^{-3}$ to $1$.
We see the system crossing the $50\%$ barrier and attaining higher excitations at higher couplings.
Figure \ref{fig2}(e) shows $P_{\mathrm{exc}}(t=10)$ as a function of $T$ at $\kappa^{(1)}=10^{-5}$.
Because the system has effectively only two energy levels in the case $\kappa^{(1)}=10^{-5}$, the maximal of $\partial_T P_{\mathrm{exc}}(t=10)$ is around $T_\theta \sim 1$, near the peak of the specific heat \cite{thirumalai1995}.
Figure \ref{fig2}(f) shows $P_{\mathrm{exc}}(t=10)$ as a function of $\kappa^{(1)}$ at $T=10$.
The higher chemical-like excitation in Fig.\ref{fig2}(f) as compared to the thermal one in Fig.\ref{fig2}(e) remarkably resemble experimental results in Ref.\cite{narayan}.

We have finally considered the scaling of pathways suppression for $N$ spins-$1/2$ with nondegenerate gaps.
As in Eq.(\ref{HSR}), each spin is coupled to an independent reservoir. 
We find that the minimum number of zeros $N_{\mathrm{zeros}}$ in $\Lambda$ is given by
$N_{\mathrm{zeros}} = 2^N \left[2^N - (N+1)\right]$.
This can be understood by noticing that each line in $\Lambda$ contains $N+1$ nonzero elements.
Hence, $2^N - (N+1)$ zeros.
The number of lines is $2^N$, explaining the result in $N_{\mathrm{zeros}}$.
For large chains, $N \gg 1$, the number of zeros approaches the number of matrix elements, $2^{2N}$.
The almost linear growth of the number of allowed relaxation pathways is, therefore, unable to ensue the exponential growth of the system dimension.
This behavior qualitatively reminds us of protein physics, in the sense that the amount of blocked relaxation pathways are typically much larger than the allowed ones for bigger chains.

Before concluding, we would like to state what we believe to be the fundamental link between our general formalism and more realistic models of thermal and chemical denaturation in proteins.
The molecular transfer model \cite{thirumalai2008,thirumalai2010, thirumalai2011,reviewJE,thirumalai2016} combines coarse-grained molecular dynamics simulations and Tanford's transfer model \cite{tanford64} to accurately predict the dependence of equilibrium properties of proteins at finite concentration of osmolytes and denaturants.
Tanford's model distinguishes those peptide groups that are in contact with the surrounding environment (solvent-accessible surface area) from those that are, by contrast, shielded from the solvent by other parts of the protein molecule.
This asymmetric coupling to the environment consists in the working principle of his model to capture chemical denaturation.
Here, we have addressed, from a completely different approach, a generalization of that asymmetric environment idea.
It turns out that such a generalization implies resemblances between radically distinct systems.
It may be the case that our results provide a more fundamental connection between the (stochastic) Langevin approach for describing thermal effects and the (thermodynamic) transfer model for describing chemical effects \cite{thirumalai2011, thirumalai2016} by means of bead-dependent and state-dependent friction coefficients and random forces.
It may also be the case that the ion jacket picture that explains polyelectrolyte conformations \cite{gold2019}, where the spatial distribution of surrounding ions matters as much as their concentration to the polymer's shape, consists in another realization of our general asymmetric environment theory.


%

In conclusion, we have shown a general framework for protein-like relaxation dynamics to emerge from asymmetric couplings between a spin chain and its environment.
A strongly asymmetric environment can block the system from finding its minimum energy state, complementing the role of frustration in protein folding.
Remarkably, a strongly asymmetric environment can also induce distinct thermal and chemical-like excitation pathways in a spin chain, as reminiscent of protein unfolding dynamics.
Our results open a research line in which reservoirs and relaxation pathways can be devised, along with system's Hamiltonians (including time-dependent drives \cite{JE15}), intended to make emerge from inorganic chains all the other typical dynamical aspects of proteins, e.g. the iterative annealing mechanism of chaperonins and allostery, contributing to our understanding of life-like behavior.


\begin{acknowledgements}
We thank P. H. L. Martins for useful comments.
D. V. and T. W. acknowledge support from INCT-IQ, Brazil.
\end{acknowledgements}


%

%

%

%

%
\end{document}